# ΔSCF Excitation Energies Up a Ladder of Ground-State Density Functionals

Ethan Pollack, Rohan Maniar, and John P. Perdew*
Department of Physics and Engineering Physics,
Tulane University, New Orleans, Louisiana, USA 70118

Density functional theory (DFT) is a widespread and effective tool in electronic structure calculations for ground-state electron systems. Its success has prompted exploration into the use of DFT for non-collective excited-states. The delta self-consistent field (ΔSCF) method allows for the extension of DFT to excited-state energies by restricting the Kohn-Sham orbital occupations, producing an excited-state electron density, and then computing its energy. In this paper, we examine the performance of the LSDA, PBE generalized-gradient approximation (GGA), and SCAN/r$^2$SCAN meta-GGA for the excitation energies of several important systems. We consider the energies of atoms with atomic number 1-18. For the hydrogen atom, where we use the exact electron density and have no multiplet splitting, we find significant improvement up the ladder from LSDA to PBE to SCAN. For the uniform gas, we find an effective mass different from the bare mass only with r$^2$SCAN. We split the case of multi-electron atoms into non-aufbau excitations, where the highest-energy electron is excited to the lowest state in the next $nl$ subshell (where accuracy is least limited by available basis sets), and spin-flip excitations, where the spin of an electron is flipped, leading to a higher-energy state of the same $nl$ configuration. We find reasonably accurate approximate excitation energies, except for the spin-flip cases where the auxiliary non-interacting wavefunction of a non-Hund's-rule spin state is not well described by a single-determinant.

## Introduction

Kohn-Sham density functional[1] (KS-DFT) has had remarkable success in predicting the ground-state properties of material systems, having become a go-to method for the calculation of energies for single atoms, molecules, and condensed matter. Since excited states are also of wide interest and utility, it would be important to be able to apply advances in functional approximation to the calculation of excitation energies from total-energy differences.

The constrained search formulation[2] of DFT provides an expression for the exact ground-state functional, which has enabled the derivation of many of its mathematical properties that can constrain computationally-feasible approximate functionals for the exchange-correlation energy. Given the exact density functional for the exchange-correlation energy, Kohn-Sham DFT would be exact for the ground-state total energy and electron density for electrons in the presence of a scalar external potential.

Recently, there has been exploration into the possibility that DFT is exact for excited states as well, motivated by the delta self-consistent field (ΔSCF) method for calculating excited-state energies. This method involves restricting the Kohn-Sham orbital occupations by requiring electrons to fill excited spin orbitals, rather than allowing the ground-state filling, and calculating the energy for the associated density. Yang and Ayers[3] suggested in their recent work that ground and excited states require the same universal functional of the density and occupied orbitals, which could explain why ground-state functionals often successfully describe excited-state energies in ΔSCF calculations[4,5], which fall outside their intended use.

SCAN[6] and r$^2$SCAN[7] are meta-generalized gradient approximation (mGGA) density functionals satisfying all 17 known exact constraints on the exact functional that a meta-GGA can. In this paper, we investigate r$^2$SCAN's performance with excited states to determine the ability of meta-GGAs to handle excited states. Where appropriate, we compare with the local spin-density approximation (LSDA) and PBE generalized-gradient approximation (GGA), which sit on lower rungs of the Jacob's Ladder[8] of density functional approximations, to examine whether increased accuracy for ground state densities extends universally to all densities. While LSDA and PBE are explicit functionals of the density, for which the effective exchange-correlation potential is a multiplication operator, the SCAN-like functionals depend explicitly on the density and occupied orbitals, implemented in a generalized Kohn-Sham scheme in which that effective potential is an integro-differential operator.

We investigate two extreme limits: free atoms and the uniform electron gas. First, we compare the excitation energy predictions of the approximate functionals in the hydrogen atom to the known actual energies. The hydrogen atom excitations provide a window into r$^2$SCAN's ability to handle neutral one-electron excitations. Then, we examine its performance predicting the separated electron-hole pair effective mass of an excitation in the UEG, a longstanding problem. In contrast to the hydrogen atom, this is a many-electron system, which provides information about r$^2$SCAN's handling of excitations when the electron-electron interaction becomes important. Finally, we compute excitation energies for valence electrons in elements with atomic numbers 2-18, evaluating r$^2$SCAN's performance in experimentally-attainable, multi-electron systems.

*Corresponding Author perdew@tulane.edu

## Computational Details

We performed all calculations using the UTEP-NRLMOL-based FLOSIC code[9,10], which has an accurate numerical integration grid and basis set for ground-state calculations. For the excited-state calculations, we used the default NRLMOL basis set in addition to diffuse even-tempered Gaussian functions of s-, p-, and d-type. All calculations were performed to a tolerance of $10^{-8}$ Hartree. We constrained excited-state calculations to integer occupations to enforce convergences to excited-state densities. For non-aufbau excitations, we also enforced maximum $M_s$, consistent with our experimental energy and state references from NIST[11].

## Results

### Excitations in the Hydrogen Atom

First, we probe r$^2$SCAN's predictions of the excited states in the hydrogen atom. The exact energy of the hydrogen atom is given by

$$E_{Total} = T_s[n] + \int d^3\vec{r}\left(-\frac{e^2}{r}\right)n(\vec{r}) + U[n] + E_{xc}[n]. \quad (1)$$

For any one-electron density

$$E_{xc}^{Exact} = -U[n], \quad (2)$$

where $U[n]$ is the Hartree electrostatic energy. Since the exact stationary state densities are known, we can use them to compute the exact exchange-correlation energies from Eq. (2) as well as from our density functional approximations. From Table 2 in Sun et al. [12], we produce Table 1, which contains the exact exchange-correlation energies and the predictions from LSDA, PBE, and SCAN.

The exact total energy for any bound state is known analytically:

$$E_{Exact}^{Hydrogen} = -\frac{13.6\ [eV]}{n^2} \quad (3)$$

Because approximate density functionals do not give the exact exchange-correlation, we can determine the difference between the predicted and exact exchange-correlation energy and add it to the exact hydrogen atom state energy to determine the functional prediction for state energies. This is done using the exact and analytically-known densities. Then, taking the difference of an excited state and the ground state will yield the energy required to move a ground-state electron into the chosen excited state, as

$$E_{Excitation}^{Approx} = E_{Excited\ State}^{Approx} - E_{Ground\ State}^{Approx} \quad (4)$$

These results are displayed in Table 2.

The predictions from SCAN will match the predictions from r$^2$SCAN for any single-orbital density, and both will be exact for the hydrogen atom ground-state, as it is one of the appropriate norms used in the construction for SCAN, a feature which is unaltered by r$^2$SCAN.

We pause to note that the Perdew-Zunger self-interaction correction[13] is exact for both the ground and the excited states of the one electron atom and all one-electron atomic ions. Since the corrected functional depends on the electron density through the occupied orbitals, this is a simple but interesting illustration of the Yang-Ayers conclusion.

As shown in Table 1, the error in the exchange-correlation energy from SCAN is significantly smaller than the error from earlier functionals. Also of note is the worsening performance from LSDA to PBE. This can be attributed to LSDA's better performance for noded[14] ($l \neq 0$) states, which compensates in the average error measures for LSDA's significantly reduced accuracy for the ground state. In contrast to PBE, however, SCAN outperforms LSDA for all radially-noded states except for the (2,1,0) state.

The error in the excitation energies is at least halved from LSDA to PBE and again from PBE to SCAN. This can be attributed to SCAN's improved performance over LSDA and PBE for excited states, combined with its exact reproduction of the ground state energy. On the other hand, the excitation energies for PBE are far more accurate on average than those of LSDA because its prediction for the ground state energy is far more accurate than the prediction of LSDA. Since the magnitude of the ground state energy is more than four times larger than the next largest magnitude, this prediction has an outsized impact on the accuracy of the excitation energies.

a)

| $(n, l, m)$ | $E_{xc}^{Exact}$ | $E_{xc}^{LSDA}$ | $E_{xc}^{PBE}$ | $E_{xc}^{SCAN}$ |
|---|---|---|---|---|
| (1 0 0) | −8.5036 | −7.8998 | −8.4866 | −8.5036 |
| (2 0 0) | −2.0463 | −2.1732 | −2.3471 | −2.1629 |
| (2 1 0) | −2.6626 | −2.8570 | −3.0567 | −2.8969 |
| (3 0 0) | −0.9034 | −1.0371 | −1.1211 | −0.9766 |
| (3 1 0) | −1.0561 | −1.2842 | −1.3845 | −1.2293 |
| (3 2 0) | −1.2542 | −1.4799 | −1.5928 | −1.4310 |
| (4 0 0) | −0.5072 | −0.6148 | −0.6650 | −0.5549 |
| (4 1 0) | −0.5731 | −0.7438 | −0.8034 | −0.6825 |
| (4 2 0) | −0.6210 | −0.8159 | −0.8849 | −0.7526 |
| (4 3 0) | −0.7293 | −0.9189 | −0.9940 | −0.8554 |

b)

| | $\Delta E_{LSDA}$ | $\Delta E_{PBE}$ | $\Delta E_{SCAN}$ |
|---|---|---|---|
| MAE | 0.2175 | 0.2513 | 0.1189 |
| MARE | 18.34 | 26.20 | 12.01 |

**Table 1:** Hydrogen atom exchange-correlation energies (a) Exact exchange-correlation energies and predictions from LSDA, PBE, and SCAN functionals evaluated on the exact densities. Energies in eV. (b) Mean absolute error (MAE) and mean absolute relative error (MARE) for functional predictions, with MAE in eV and MARE in %.

a)

| $(n, l, m)$ | $\Delta E^{Exact}$ | $\Delta E^{LSDA}$ | $\Delta E^{PBE}$ | $\Delta E^{SCAN}$ |
|---|---|---|---|---|
| (2 0 0) | 10.20 | 9.469 | 9.882 | 10.08 |
| (2 1 0) | 10.20 | 9.402 | 9.789 | 9.966 |
| (3 0 0) | 12.08 | 11.35 | 11.85 | 12.02 |
| (3 1 0) | 12.08 | 11.26 | 11.74 | 11.92 |
| (3 2 0) | 12.08 | 11.26 | 11.73 | 11.91 |
| (4 0 0) | 12.75 | 12.04 | 12.58 | 12.70 |
| (4 1 0) | 12.75 | 11.98 | 12.50 | 12.64 |
| (4 2 0) | 12.75 | 11.95 | 12.47 | 12.62 |
| (4 3 0) | 12.75 | 11.96 | 12.47 | 12.62 |

b)

| | $\Delta E^{LSDA}$ | $\Delta E^{PBE}$ | $\Delta E^{SCAN}$ |
|---|---|---|---|
| MAE | 0.7246 | 0.2806 | 0.1321 |
| MARE | 6.552 | 2.512 | 1.133 |

**Table 2:** Hydrogen atom excitation energies energies (a) Exact excitation energies and predictions from LSDA, PBE, and SCAN functionals evaluated on the exact densities. Energies in eV. (b) Mean absolute error (MAE) and mean absolute relative error (MARE) for functional predictions, with MAE in eV and MARE in %.

The carbon dimer provides a similar example of SCAN's improved performance in excited states over PBE and LSDA. Perdew et al. found that SCAN is the only functional of the three which correctly predicts a positive excitation energy from the strongly-correlated singlet ground state of $C_2$ to its low-lying triplet excited state, and that it does so only after symmetry breaking[15]. A reference quantum Monte Carlo calculation found an excitation energy of 0.25 eV[16]. Of the three tested functionals, only SCAN was able to predict a positive excitation energy, which it found to be 0.19 eV. This agrees with our calculated MAE of ~0.13 eV, confirming that SCAN can differentiate between states with energy differences on the order of 0.1 eV.

## Excitation in the Uniform Electron Gas

Here we consider a model system of real interacting electrons neutralized by a rigid uniform positive charge density. The electron density is $n = 3/[4\pi(r_s a_0)^3]$, where $a_0 = \frac{\hbar^2}{me^2}$, the uniform electron gas (UEG). This system, which roughly models valence-electron regions in real metals ($1 \leq r_s \leq 6$), has been intensely studied in many-body physics[17]. Over a wide range of the density parameter $r_s$, the UEG is a Fermi liquid with a discontinuity in its natural occupation numbers at the Fermi level (with the discontinuity increasing to 1 as $r_s \to 0$). Its neutral excitations include those corresponding to two quasiparticles, a delocalized hole below the Fermi level and a delocalized electron above it, When both one-electron states are close to the Fermi level, the excitation energy is renormalized from its non-interacting value by a change of effective mass. Since the ground-state and the excited-state electron densities are both uniform and both the same, an exchange-correlation functional of the electron density alone (e.g., LSDA, PBE, or the original Kohn-Sham exact density functional) will predict the same excitation energies for the interacting as for the non-interacting systems, and so cannot correctly describe its quasiparticle excitation energies. But, since the occupied orbitals change upon excitation, an orbital-dependent functional (like a meta-GGA, a hybrid with exact exchange, or a self-interaction correction) has a chance to do so.

We now turn our attention to r²SCAN's prediction of the electron-hole quasiparticle effective mass in the UEG (since SCAN, like the aforementioned functionals, gives $m^*/m = 1$ for all $r_s$). We calculate the total change in energy due to the excitation as a function of the density parameter $r_s$, to determine the ratio of the effective to the bare electron mass, a ratio that is 1 for non-interacting electrons. SCAN and r²SCAN meta-GGAs use the kinetic energy density $\tau$, as well as $n$ and $\nabla n$, to interpolate between different GGAs for single-orbital and uniform density regions, and to extrapolate to regions of density-tail overlap. We focus only on r²SCAN in this section.

A detailed derivation of the quasiparticle effective mass is provided in S2. The key terms and their relevance are described here. The kinetic energy density appears in r²SCAN through the dimensionless iso-orbital indicator

$$\bar{\alpha} = \frac{\tau - \tau_W}{\tau_{unif} - \eta\tau_W} \quad (5)$$

Here, $\tau$ is the non-interacting kinetic energy density of the system (constructed from the occupied orbitals), $\tau_W = \frac{\hbar^2}{m}\frac{|\nabla n^2|}{8n}$ is the von Weizsäcker kinetic energy density, $\tau_{unif} = \frac{3}{10}\frac{\hbar^2}{m}(3\pi^2)^{2/3}n^{5/3}$ is the kinetic energy density of the uniform electron gas, $n$ is the electron density, and $\eta = 10^{-3}$ is a regularization parameter. For the UEG, $\tau_W = 0$, so $\bar{\alpha} = 1$. Now, we introduce an excitation to an arbitrary electron below the Fermi level to a state above it, so that $|\boldsymbol{k}| < |\boldsymbol{k} + \boldsymbol{q}|$, where $\boldsymbol{k}$ is initial the wavevector of the arbitrary electron with kinetic energy $\frac{\hbar^2}{2m}|\boldsymbol{k}|^2$ and $\boldsymbol{q}$ is the change in the wavevector from the excitation. This leads to a change in the kinetic energy density

$$\tau \to \tau + \Delta\tau \quad (6)$$

$$\Delta\tau = \frac{\Delta E_{KS}}{V} \quad (7)$$

where $V$ is the volume and $\Delta E_{KS} = \frac{\hbar^2}{2m}[2\boldsymbol{k}\cdot\boldsymbol{q} + \boldsymbol{q}^2]$ is the excitation energy of the Kohn-Sham auxiliary non-interacting system. This leads to corrections to the exchange and correlation energies shown in Eqs. (S15) and (S20) of the Supporting Information.

In our Eq. (10) below, all parameters come from r²SCAN[6] except those for the correlation energy of the uniform electron gas, for which we use the simpler Ceperley-Alder form[18], based on quantum Monte Carlo simulations, which is highly accurate for $1 \le r_s \le 100$. In the range $r_s < 1$, this form is less accurate, but because the exchange energy dominates in the limit $r_s \to 0$ this should not significantly impact the effective mass ratio. Furthermore, as $r_s \to 0$, our effective mass ratio correctly approaches 1.

The total energy of the neutral excitation is the sum of the excitation energy of the Kohn-Sham auxiliary system and the change of the exchange-correlation energy.

$$\Delta E_{total} = \Delta E_{KS} + \Delta E_{xc} = \Delta E_{KS}\left[1 + \frac{\Delta E_{xc}}{\Delta E_{KS}}\right] \quad (8)$$

We can now include the masses explicitly and solve for the effective mass ratio from the total change in energy. Let $a_1 = \left(\frac{1}{12\pi^2}\right)^{\frac{2}{3}}$, $a_2 = \left(\frac{16}{2187\pi^2}\right)^{\frac{1}{3}}$ for compactness. Then, in terms of $r_s$, we can write the effective mass ratio as

$$\Delta E_{Total} = \frac{\hbar^2}{2m}\left[2\vec{k}\cdot\vec{q} + \vec{q}^2\right]\left\{1 + \frac{\Delta E_{xc}}{\Delta E_{KS}}\right\} = \frac{\hbar^2}{2m^*}\left[2\vec{k}\cdot\vec{q} + \vec{q}^2\right] \to \frac{m^*}{m} = \left\{1 + \frac{\Delta E_{xc}}{\Delta E_{KS}}\right\}^{-1} \quad (9)$$

$$\frac{m^*}{m} = \left\{1 - 10\left[a_1 k_0 r_s \sum_{i=0}^{7} i\cdot c_{xi} + a_2 {r_s}^2\left[\frac{-b_{1c}}{1 + b_{2c}\sqrt{r_s} + b_{3c}r_s} - \frac{\gamma}{1 + \beta_1^U\sqrt{r_s} + \beta_2^U r_s}\right]\cdot\sum_{j=0}^{7} c_{cj}\cdot j\right]\right\}^{-1} \quad (10)$$

**Figure 1:** Effective mass ratio of a quasiparticle excitation in the UEG, decreasing monotonically with density parameter $r_s$ for $0 \leq r_s \leq 10$

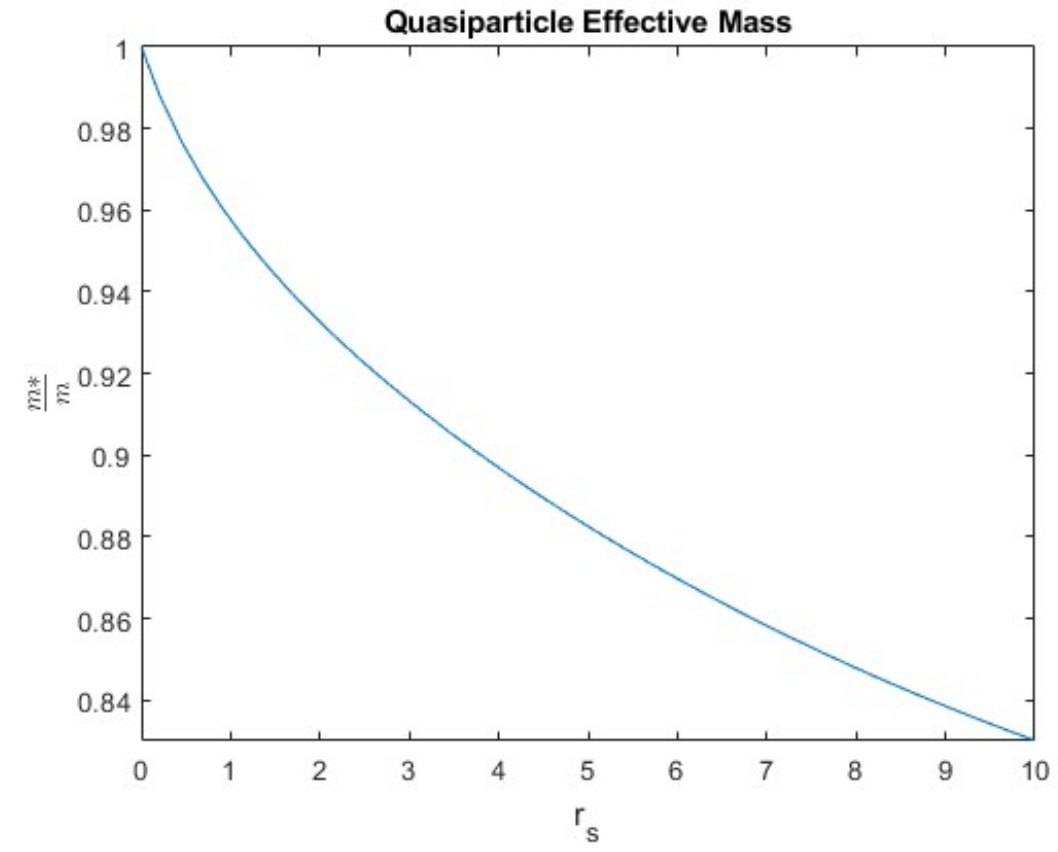

Figure 1 displays the ratio in the range $0 \leq r_s \leq 10$, and shows a monotonically decreasing ratio. This range covers the metallic solids. The quasiparticle effective mass is a point of ongoing investigation. Figure 8.22 in Giuliani and Vignale[17] displays different predictions for the effective mass ratio calculated via spectral analysis. Whether the effective mass is smaller or larger than the bare mass in the metallic range according to their calculation depends on a term which can be either positive or negative. They state that it is more likely $m^* > m$ but inconclusive. Several results, most notably one using the Fermi Liquid model, found a monotonically decreasing quasiparticle mass ratio in the metallic range[19,20]. However, a minimum value of $r_s$ in the metallic range has significant support[21,22,23]. Most notably, diagrammatic quantum Monte Carlo (DQMC) calculations found a minimum value less of about 0.94 at $r_s$ between 2 and 3[21]. While the curve in Fig. 1 shows no minimum in the plotted range, it does take a minimum around $r_s = 23.6$.

As shown, r$^2$SCAN predicts that the effective mass ratio is monotonically decreasing for metallic densities. The prediction made by r$^2$SCAN is similar in magnitude to a number of previous predictions. It also correctly predicts that the exchange-correlation correction goes to 0 in the high-density limit, where the kinetic energy dominates. Similarly, it captures the exchange dominance over correlation in the metallic range, giving us a decreasing effective mass ratio via equation (10). The graphs of the exchange, correlation, and exchange-correlation energy corrections are given in figure S1. The presence of these markers of accuracy suggests the ability of orbital-dependent functionals to handle one-particle-like excitations.

**Excitations in Multi-Electron Atoms**

The encouraging nature of the results from excitations to the hydrogen atom and the UEG raise the question of how r$^2$SCAN performs in real systems with significant complexity. To begin exploring such systems, we evaluated the performance of r$^2$SCAN in selected electron excitation energies from energy differences as

$$E_{excitation} = E_{excited} - E_{gs} \qquad (11)$$

We provide values for atoms with atomic numbers $Z$ ranging from 2 (He) to 18 (Ar), again comparing to LSDA and PBE. Here we find the electron densities from self-consistent calculations for each approximate functional. To minimize the errors of Gaussian basis sets optimized for the ground state, we focus only on low-energy excitations.

We separate these excitations into two categories, non-Aufbau and spin-flip. Non-Aufbau excitations refer to those excitations which raise the highest-energy electron to the next $nl$ subshell, while spin-flip excitations refer to those which change the spin of one electron without changing the $nl$ subshell it occupies. Note that the energy remains unchanged if we flip one spin in an $nl$ subshell where the number of electrons of one spin differs by 1 from the number of electrons of the other spin. Thus, a spin-flip excitation is only possible in a ground-state atom where the highest $nl$ subshell with electron occupation has at least two more electrons in one spin channel than the other and results in a decreased total spin. This occurs for atomic numbers 6, 7, 8, 14, 15, and 16. The two excitation types are neatly visualized in figure 2. The mean absolute error (MAE) and mean absolute relative error (MARE) for both excitation types are then listed in Tables 3 and 4, while the energy predictions for each excited state are given in S3 and S4.

**Figure 2:** Non-aufbau and spin-flip excitations in the carbon atom ($Z = 6$). Arrows denote electron spin in each column. Dashed lines separate spin channels while solid vertical lines separate $nl$ configurations.

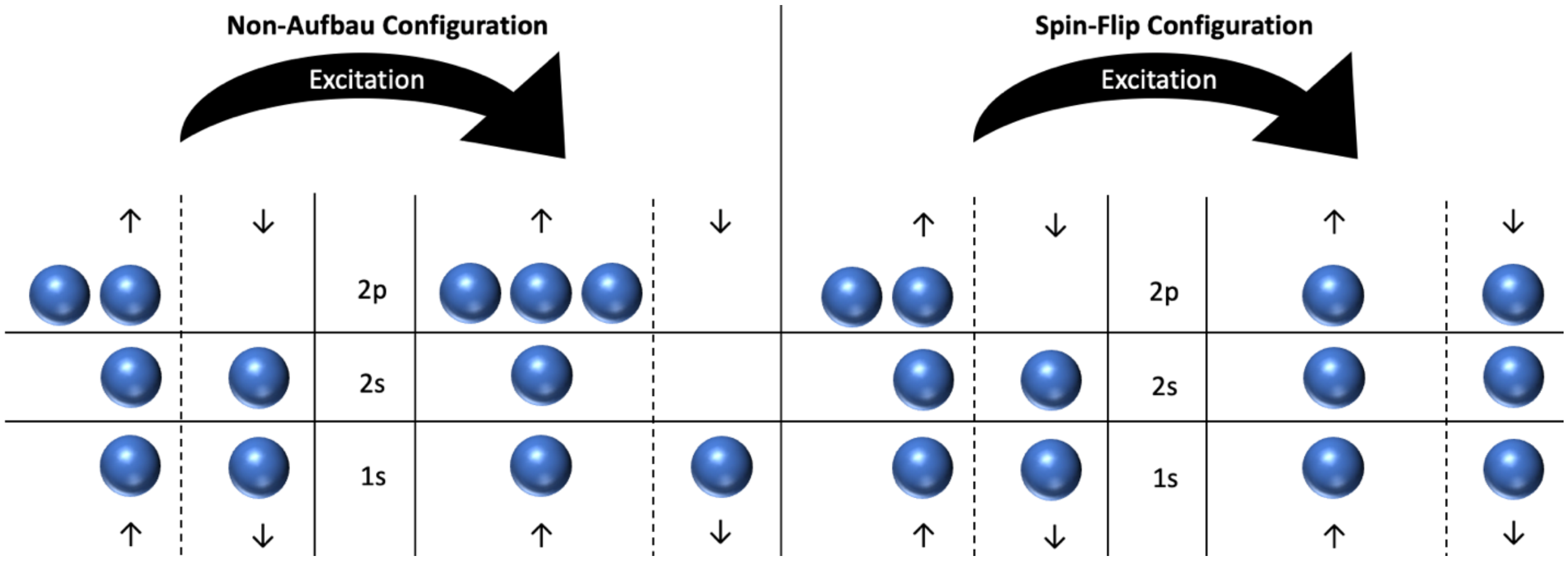

| Functional | MAE | MARE |
|---|---|---|
| LSDA | 0.20 | 3.49 |
| PBE | 0.15 | 3.37 |
| $r^2$SCAN | 0.27 | 6.38 |

**Table 3:** MAE and MARE for non-aufbau excitation energies of atoms with $Z$ from 2 to 18. MAE in eV, MARE in %

| Functional | MAE | MARE |
|---|---|---|
| LSDA | 0.69 | 48.8 |
| PBE | 0.89 | 60.5 |
| $r^2$SCAN | 0.87 | 59.2 |

**Table 4:** MAE and MARE for spin-flip excitation energies of atoms with $Z$ from 2 to 18. MAE in eV, MARE in %

In the non-aufbau case, we find a different trend from that for the hydrogen atom: PBE improves upon LSDA, but $r^2$SCAN is less accurate than either. LSDA manages to improve its errors compared to hydrogen, while each of PBE and $r^2$SCAN become relatively less accurate. However, the relative similarity of the MAREs between the one-electron and many-electron cases, and the good PBE results of Ref. 5, suggest that excitations to the lowest energy state in an unoccupied *nl* subshell can be described accurately by ground-state functionals.

Reference 5 makes an extensive ΔSCF study of non-aufbau excitations in multi-electron atoms using a variety of functionals (but not LSDA or $r^2$SCAN). It distinguishes between valence excitations (in which the transitioning electron does not change its principal quantum number $n$, as in 7 of our 17) and Rydberg excitations (in which it does, as in 10 of our 17). Section V of Ref. 5 discusses the basis-set convergence for each kind of non-aufbau excitation. As in our study, Ref. 5 considers only transitions in which the initial and final spin states are both appropriate for density functional calculations with a single Slater determinant.

$r^2$SCAN performed significantly better than both LSDA and PBE in the case of the single-orbital hydrogen atom, where the exact densities for the excited states were used to compute the energies. The tables in sections 3 and 4 of the Supplementary Information show that $r^2$SCAN also performs well for all but 5 of the 17 non-aufbau excitations. Its large errors arise only for $Z$ = 4, 5, 6, 12, and 14, in which an electron transitions from $(ns)^2$ to $nsnp$, creating regions of strong spin-polarization in which different orbital shapes are strongly overlapped. This is a region that is not well controlled by $r^2$SCAN's appropriate norms, suggesting the possibility of improving $r^2$SCAN without losing any of its exact constraints and appropriate norms.

The default NRLMOL ground-state basis set is clearly insufficient for the first non-aufbau excited state, yielding MAEs about 5 times bigger than the ones we report here that augment each default s, p, or d Gaussian basis function by one similar but diffuse Gaussian.

In contrast to the non-aufbau cases, each functional sees a large MARE on the order of 50% in the spin-flip cases. Of the three, LSDA performs best. Ref. 5 also observed that ground-state functionals perform much better for non-aufbau than for spin-flip excitations. We speculate that the 50% underestimation of the spin-flip excitation energy arises because our auxiliary non-interacting wavefunction for the excited state, a single Slater determinant with $M_s = S_{max} - 1$, is roughly a linear combination of $S_{max} - 1$ and $S_{max}$ contributions.

A standard Kohn-Sham calculation for an atom employs integer occupation number for a set of spin orbitals, and chooses the maximum possible $M_s$. This is fine as long as one is searching for the lowest-energy term of a multiplet, which by Hund's rule must have the maximum possible total spin $S$. That is the right term for a ground state, or for the excited state in a non-aufbau excitation of lowest excitation energy. But it is not even possible for the final state of a spin-flip excitation, which may explain the large relative error in Table 4. At least the predicted excitation energies have the correct sign. The case of spin flips is a remaining challenge for DFT.

One notable example of accurate excitation energies comes from Wang and Chong[4], who found a 0.07 eV mean absolute deviation (MAD) of 59 best cases (of 68 total) vertical carbon 1s binding energies in non-polar or weakly polar molecules using a ground-state exchange-correlation functional with no orbital dependence. For highly polar molecules, only hybrid functionals with orbital dependence gave accurate excitation energies, which is consistent with the need for orbital dependence to describe ground-state energy differences in the presence of strong electron transfer between atoms, and with Ref. 3.

## Conclusions

Yang and Ayers[3] have suggested that the same exchange-correlation functional of the density and occupied orbitals can correctly describe ground and excited states of electronic systems. In this work, we have tested non-empirical ground-state approximations (LSDA, PBE, and r$^2$SCAN) on the first three rungs of Jacob's Ladder[7] for their ability to predict excitation energies.

For one-electron systems, the exact orbital functional[13] is known to predict the exact excited-state densities and energies. For the hydrogen atom, the more computationally efficient functionals were tested here on the known exact densities, with a good accuracy that increases up the ladder, confirming that self-interaction error decreases up the ladder.

For non-collective excitation energies of the uniform electron gas, functionals that depend only on the excited state density (LSDA, PBE, and even the exact density functional of Kohn-Sham theory) are not always inaccurate, but they cannot predict a many-body variation of quasiparticle effective mass with density parameter $r_s$, which only orbital-dependent functionals including r$^2$SCAN can do.

For multi-electron atoms, the ground-state functionals are reasonably accurate for excitation energies from the ground state to of lowest-energy excited state of a different $nl$ configuration, but the expected improvement up the ladder is not seen. The converged ground-state densities of main group molecules and closed-subshell atoms are known to improve significantly from LSDA to PBE to SCAN/r2SCAN[24-28]. illustrating the predictive power of the exact constraints and appropriate norms[29]. The failure of r$^2$SCAN for excited states with strong spin-polarization and strong orbital overlap may suggest improvements to r$^2$SCAN that do not require any sacrifice of exact constraints or any refitting to current appropriate norms. This possibility is now under study.

For spin-flip excitations that change the total spin but not the $nl$ configuration of a multi-electron atom, our numerical predictions strongly underestimate the excitation energies, and the best functional was LSDA, at the bottom of the ladder. Clearly, a better way to predict multiplet energy splittings in density functional theory is needed for these excitations.

## Acknowledgements

EP, RM, and JPP were supported by the National Science Foundation under Grant No. DMR-242675.

# Supporting Information for "ΔSCF Excitation Energies Up a Ladder of Ground-State Density Functionals"

Ethan Pollack, Rohan Maniar, and John P. Perdew*
Department of Physics and Engineering Physics,
Tulane University, New Orleans, Louisiana, USA 70118

1. Exchange, Correlation, and Exchange-Correlation Energy Corrections to a Quasiparticle in the UEG

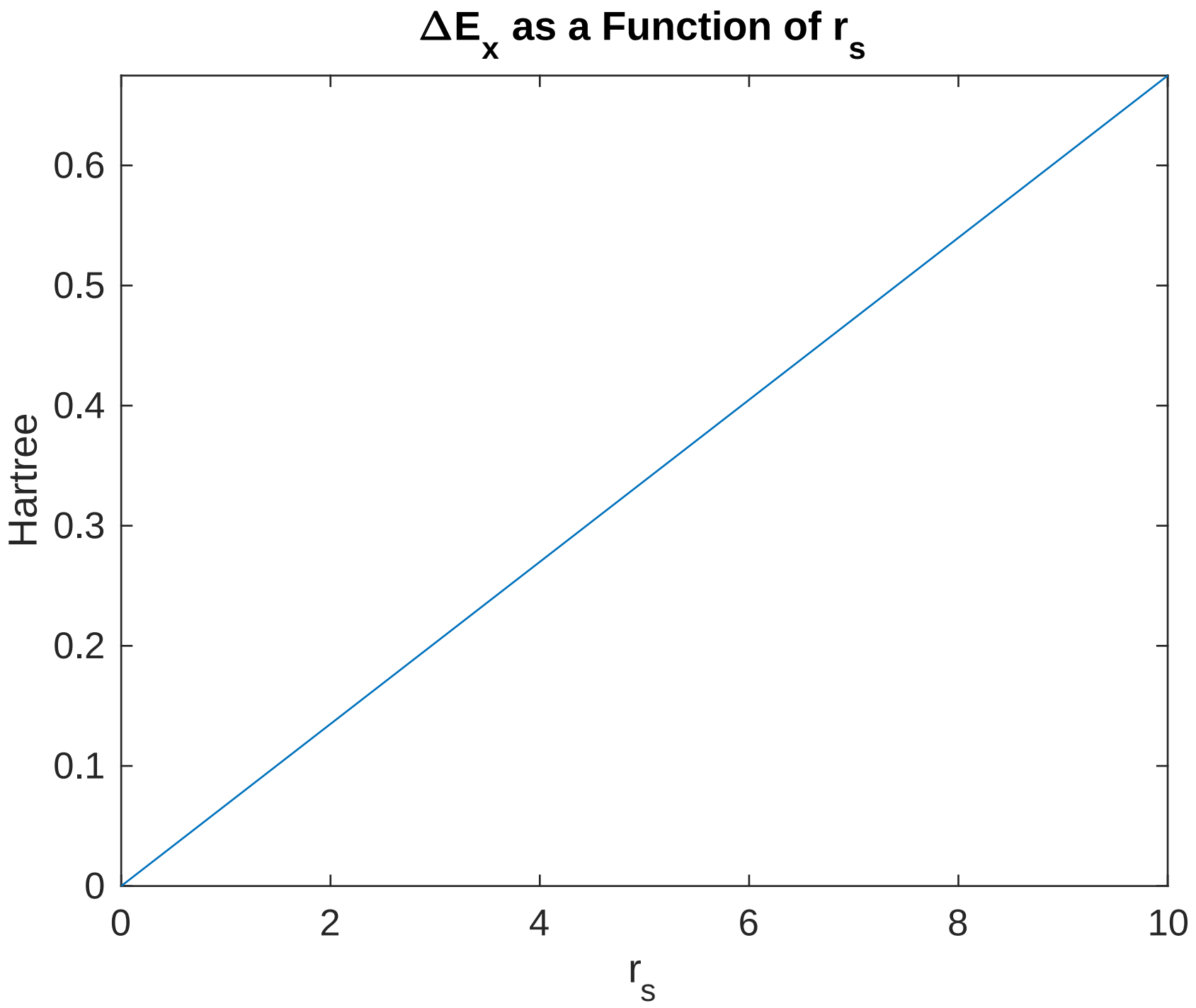

a.

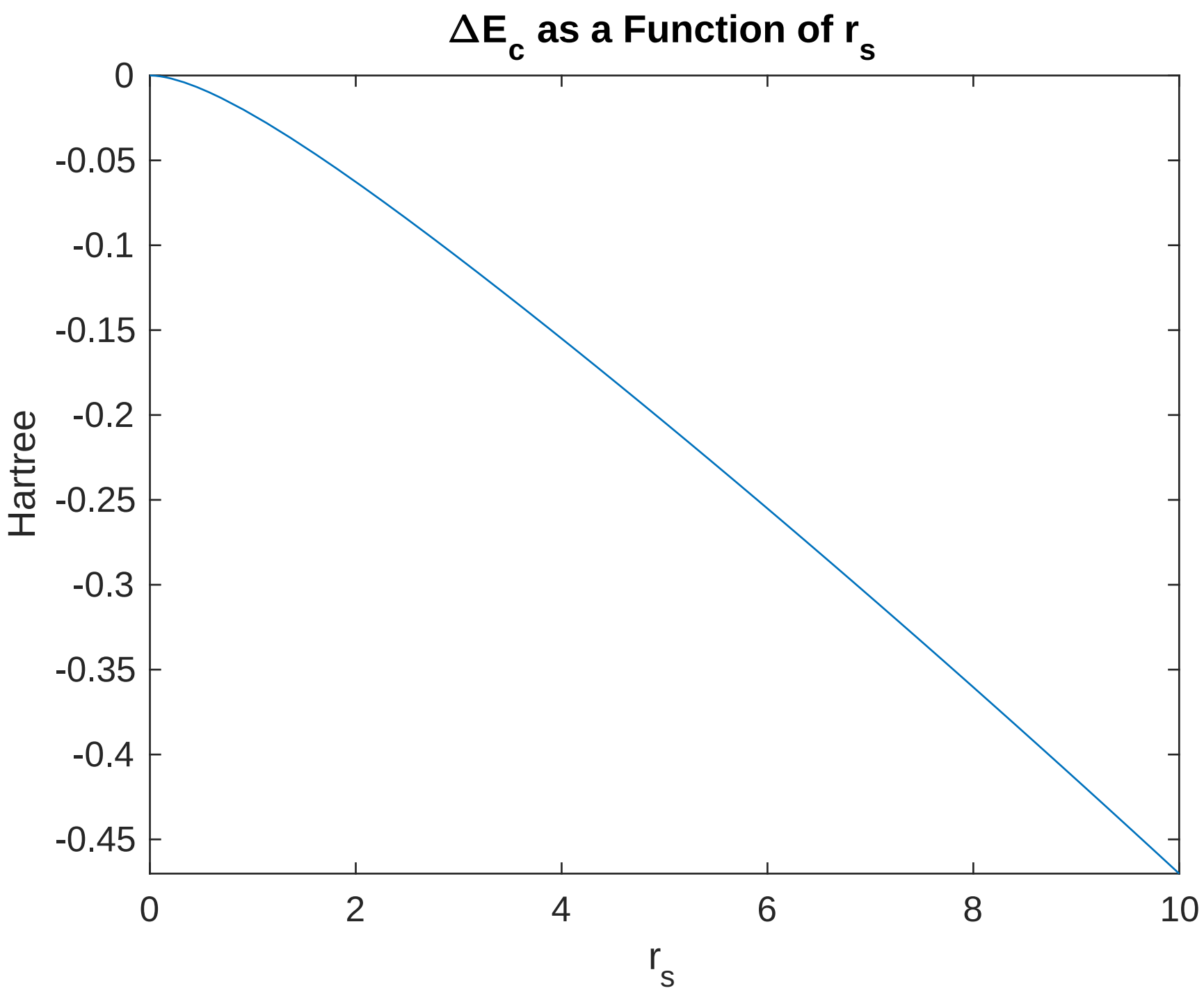

b.

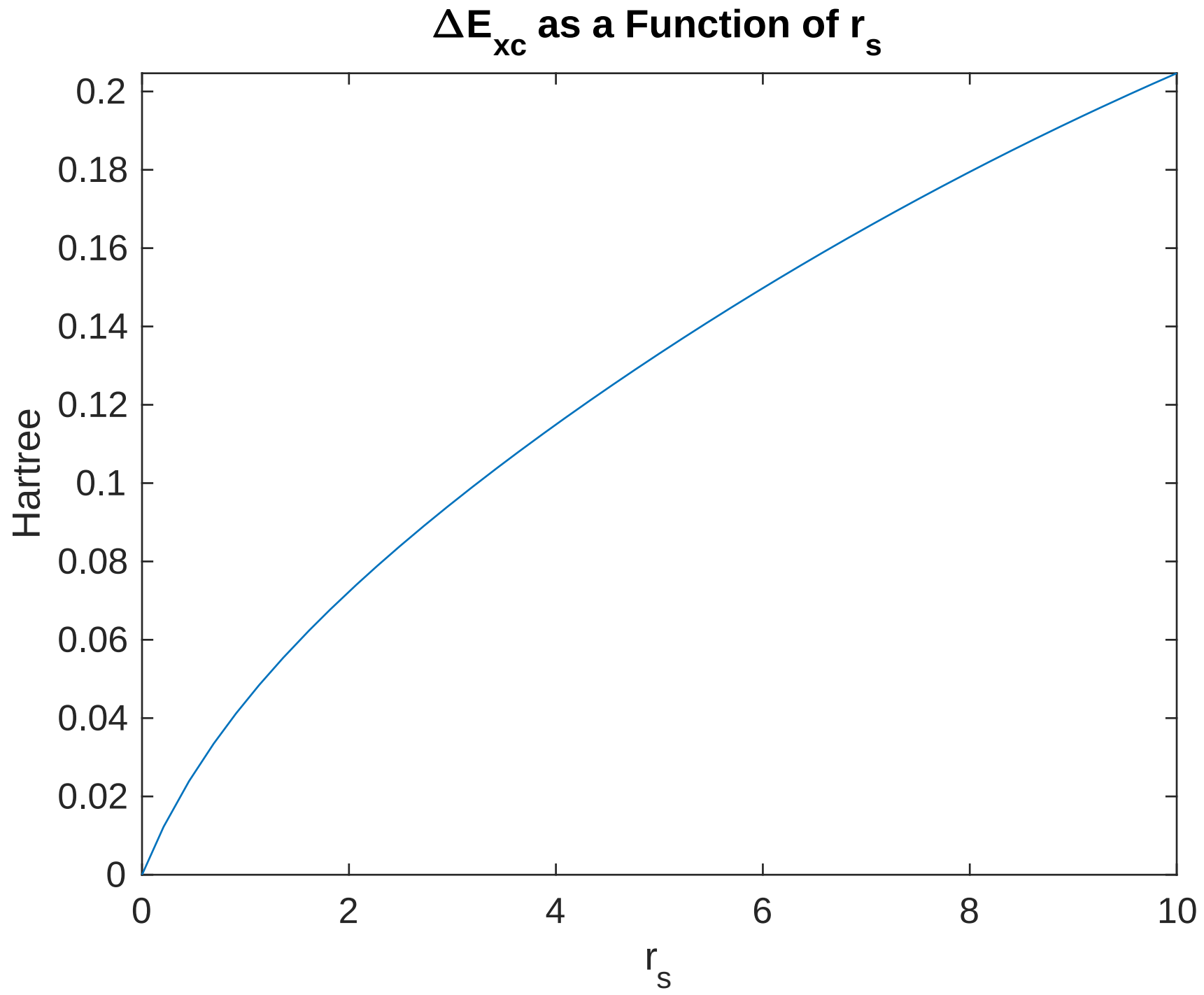

c.

## 2. Derivation of the Quasiparticle Effective Mass Ratio

We have the following definitions (in atomic units where $\hbar = m = 1$).

$$\bar{\alpha} = \frac{\tau - \tau_W}{\tau_{unif} - \eta\tau_W}, \tau_W = \frac{|\nabla n^2|}{8n}, \tau_{unif} = \frac{3}{5}(3\pi^2)^{\frac{2}{3}} n^{\frac{5}{3}}, n = \frac{k_f^3}{3\pi^2} \quad (S12)$$

For the UEG, $\nabla n = 0 \rightarrow \tau_W = 0$, so $\bar{\alpha} = 1$. Next, we apply our excitation to an electron in the UEG

$$|\boldsymbol{k}_f| < |\boldsymbol{k} + \boldsymbol{q}|, \tau \rightarrow \tau + \Delta\tau \quad (S13)$$

$$\Delta\tau = \frac{\Delta E_{KS}}{V} \quad (S14)$$

where $V$ is the volume and $\Delta E_{KS} = [2\boldsymbol{k}\cdot\boldsymbol{q} + \boldsymbol{q}^2]$ is the excitation energy to the Kohn-Sham non-interacting orbitals, leading to a correction to $\bar{\alpha}$

$$\bar{\alpha} = 1 + \frac{\Delta\tau}{\tau_{unif}} \tag{S15}$$

We assume that $\frac{\Delta\tau}{\tau_{unif}} \ll 1$ because only one electron in the gas is excited. In the range $0 \le \bar{\alpha} \le 2.5$, the interpolation function is given by

$$f(\bar{\alpha}) = \sum_{i=0}^{7} c_i\,\bar{\alpha}^i = \sum_{i=0}^{7} c_i\left[1 + \frac{\Delta\tau}{\tau_{unif}}\right]^i \tag{S16}$$

The only difference between the exchange and correlation interpolation functions $f_x$ and $f_c$ is the specific set of constants $c_i$, which are denoted by corresponding subscripts.[7] To evaluate, we Taylor expand the interpolation function to the first order in $\frac{\Delta\tau}{\tau_{unif}}$ about 0. This gives us the Taylor-expanded interpolation function

$$f^{Taylor} = \sum_{i=0}^{7} c_i\left[1 + i\frac{\Delta\tau}{\tau_{unif}}\right] \tag{S17}$$

Note that in the ground state $\Delta\tau = 0, \bar{\alpha} = 1,$ the Taylor-expanded interpolation function returns the same value as the exact interpolation function. Additionally, higher-order terms in the expansion go to 0 as $V \to \infty$ because the dependence of $\Delta\tau$ on $\frac{1}{V}$ means that only the first order term remains nonzero after a volume integral is applied. Since $\Delta\tau$ is dependent on the system volume, this expansion is valid for small $\Delta E$. As such, the change in the exchange correlation energy comes from the change in the interpolation function due to the excitation

$$\Delta f^{Taylor} = \sum_{i=0}^{7} i\cdot c_i\cdot\frac{\Delta\tau}{\tau_{unif}} \tag{S18}$$

from which we can determine the energy differences. Starting with the exchange energy, we have

$$E_x[n] = \int d^3\boldsymbol{r}(n(\boldsymbol{r})\epsilon_x^{r^2SCAN}(r_s)) \tag{S19}$$

where

$$\epsilon_x^{r^2SCAN} = \epsilon_x^{LDA}F_x^{r^2SCAN}(p,\bar{\alpha}) \tag{S20}$$

$$\epsilon_x^{LDA} = -\frac{3}{4\pi}\left(\frac{9}{4\pi}\right)^{\frac{1}{3}}\frac{1}{r_s} \tag{S21}$$

$$F_x^{r^2SCAN} = h_x^1(p) + f_x(\bar{\alpha})[h_x^0 - h_x^1(p)]g_x(p) \tag{S22}$$

Here, $p = \left(\frac{|\nabla n|}{2k_f n}\right)^2$. Since $\nabla n = 0$ for the UEG, $p = 0$. We acquire the change in energy from (S11), which has dependence on $\Delta f^{Taylor}$. Making this substitution and applying definitions from the supporting information from Furness et al.[7] gives us

$$\Delta E_x = \int d^3\boldsymbol{r}\left(n\epsilon_x^{LDA}\Delta f_x^{Taylor}k_0\right) \tag{S23}$$

Note that none of the terms depend on position $\boldsymbol{r}$ and the integral over the volume is

$$\Delta E_x = n\epsilon_x^{LDA}\Delta f_x^{Taylor}k_0 V \tag{S24}$$

$$\Delta E_x = n\epsilon_x^{LDA}k_0\sum_{i=0}^{7} i\cdot c_{xi}\cdot\frac{\Delta E_{KS}}{\tau_{unif}} \tag{S25}$$

We can substitute terms with dependence on $r_s$, giving us

$$\Delta E_x(r_s) = -\left[\frac{1}{12\pi^2}\right]^{\frac{2}{3}}\cdot 10k_0 r_s\sum_{i=0}^{7} i\cdot c_{xi}\cdot\Delta E_{KS} \tag{S26}$$

Now, we follow the same set of steps for the correlation energy.

$$E_c[n] = \int d^3\boldsymbol{r}\left(n(\boldsymbol{r})\epsilon_c^{r^2SCAN}(r_s)\right) \tag{S27}$$

Here, we apply the relevant definitions again

$$\epsilon_c^{r^2SCAN} = \epsilon_c^1 + f_c(\bar{\alpha})(\epsilon_c^0 - \epsilon_c^1) \tag{S28}$$

$$\epsilon_c^1 = \epsilon_c^{LSDA} + H_c^1 \tag{S29}$$

$$\epsilon_c^0 = (\epsilon_c^{LDA0} + H_0)G_c(\xi) \tag{S30}$$

We use the Ceperley-Alder[18] form for $\epsilon_c^{LSDA}$. Noting that the UEG is spin-unpolarized at metallic densities, we arrive at

$$\Delta E_c = 10\left(\frac{16}{2187\pi^2}\right)^{\frac{1}{3}} \cdot {r_s}^2 \left[\frac{-b_{1c}}{1 + b_{2c}\sqrt{r_s} + b_{3c}r_s} - \frac{\gamma}{1 + \beta_1^U\sqrt{r_s} + \beta_2^U r_s}\right] \cdot \sum_{i=0}^{7} c_{ci} \cdot i \cdot \Delta E_{KS} \qquad (S31)$$

The values of $b_{1c}$, $b_{2c}$, and $b_{3c}$ come from r$^2$SCAN while the values of $\gamma$, $\beta_1^U$, and $\beta_2^U$ come from the Ceperley-Alder parameterization of the correlation energy. Now we can explicitly state the energy

$$\Delta E_{total} = \Delta E_{KS} + \Delta E_{xc} = \Delta E_{KS}\left[1 + \frac{\Delta E_{xc}}{\Delta E_{KS}}\right] \qquad (S32)$$

Including the masses explicitly and solving for the effective mass ratio gives us

$$\Delta E_{Total} = \frac{\hbar^2}{2m}[2\boldsymbol{k} \cdot \boldsymbol{q} + \boldsymbol{q}^2]\left\{1 + \frac{\Delta E_{xc}}{\frac{\hbar^2}{2m}\left[2\vec{k} \cdot \vec{q} + \vec{q}^2\right]}\right\} = \frac{\hbar^2}{2m^*}[2\boldsymbol{k} \cdot \boldsymbol{q} + \boldsymbol{q}^2] \qquad (S33)$$

$$\frac{m^*}{m} = \left\{1 + \frac{\Delta E_{xc}}{\Delta E_{KS}}\right\}^{-1} \qquad (S34)$$

3. Non-Aufbau and Spin-Flip Excitation Energy Predictions of LSDA, PBE, and r$^2$SCAN for Atomic Numbers 2-18

| $Z$ | Electron Configuration (Ground → Excited) | Experimental | LSDA | PBE | r$^2$SCAN |
|---|---|---|---|---|---|
| 2 | $1s^2$ [$^1S$] → 1s2s [$^3S$] | 19.82 | 19.45 | 19.51 | 19.86 |
| 3 | $1s^22s$ [$^2S$] → $1s^22p$ [$^2P$] | 1.85 | 1.65 | 1.60 | 1.66 |
| 4 | $1s^22s^2$ [$^1S$] → $1s^22s2p$ [$^3P$] | 2.72 | 2.50 | 2.40 | 2.24 |
| 5 | $1s^22s^22p$ [$^2P$] → $1s^22s2p^2$ [$^4P$] | 3.55 | 3.28 | 3.27 | 2.94 |
| 6 | $1s^22s^22p^2$ [$^3P$] → $1s^22s2p^3$ [$^5P$] | 4.18 | 4.04 | 4.11 | 3.61 |
| 7 | $1s^22s^22p^3$ [$^4S$] → $1s^22s^22p^23s$ [$^4P$] | 10.32 | 10.72 | 10.46 | 10.73 |
| 8 | $1s^22s^22p^4$ [$^3P$] → $1s^22s^22p^33s$ [$^5S$] | 9.15 | 9.35 | 9.40 | 9.26 |
| 9 | $1s^22s^22p^5$ [$^2P$] →$1s^22s^22p^43s$ [$^4P$] | 12.70 | 13.05 | 12.76 | 12.85 |
| 10 | $1s^22s^22p^6$ [$^1S$] → $1s^22s^22p^53s$ $^2[^3/_2]$ | 16.62 | 17.07 | 16.56 | 16.60 |
| 11 | [Ne]3s [$^2S$] → [Ne]3p [$^2P$] | 2.10 | 2.02 | 1.91 | 1.89 |
| 12 | [Ne]$3s^2$ [$^1S$] → [Ne]3s3p [$^3P$] | 2.71 | 2.76 | 2.61 | 2.38 |
| 13 | [Ne]$3s^23p$ [$^2P$] → [Ne]$3s^24s$ [$^2S$] | 3.14 | 3.07 | 3.12 | 3.35 |
| 14 | [Ne]$3s^23p^2$ [$^3P$] → [Ne]$3s3p^3$ [$^5S$] | 4.13 | 4.25 | 4.20 | 3.60 |
| 15 | [Ne]$3s^23p^3$ [$^4S$] → [Ne]$3s^23p^24s$ [$^4P$] | 6.94 | 6.89 | 6.83 | 7.16 |
| 16 | [Ne]$3s^23p^4$ [$^3P$] → [Ne]$3s^23p^34s$ [$^5S$] | 6.52 | 6.49 | 6.45 | 6.62 |
| 17 | [Ne]$3s^23p^5$ [$^2P$] → [Ne]$3s^23p^44s$ [$^4P$] | 8.92 | 9.00 | 8.83 | 9.08 |
| 18 | [Ne]$3s^23p^6$ [$^1S$]→ [Ne] $3s^23p^54s$ $^2[^3/_2]$ | 11.55 | 11.70 | 11.45 | 11.72 |

**Table S1:** Non-aufbau excitation energy predictions of LSDA, PBE, and r$^2$SCAN. Column 1 gives atomic number. Column 2 gives electron configuration of ground (left) and excited (right) states electron transitions between with term symbols (Racah notation for noble gasses). Columns 3-6 give excitation energies from experimental values or functional prediction in eV. Experimental energies obtained from NIST.[9]

| $Z$ | Electron Configuration (Ground → Excited) | Experimental | LSDA | PBE | r$^2$SCAN |
|---|---|---|---|---|---|
| 6 | $1s^22s^22p^2$ [$^3P$] → $1s^22s^22p^2$ [$^1D$] | 1.26 | 0.66 | 0.40 | 0.54 |
| 7 | $1s^22s^22p^3$ [$^4S$] → $1s^22s^22p^3$ [$^2D$] | 2.38 | 1.60 | 1.10 | 1.20 |
| 8 | $1s^22s^22p^4$ [$^3P$] → $1s^22s^22p^4$ [$^1D$] | 1.97 | 0.94 | 0.72 | 0.66 |
| 14 | [Ne]$3s^23p^2$ [$^3P$] → [Ne]$3s^23p^2$ [$^1D$] | 0.78 | 0.33 | 0.29 | 0.29 |
| 15 | [Ne]$3s^23p^3$ [$^4S$] → [Ne]$3s^23p^3$ [$^2D$] | 1.41 | 0.82 | 0.70 | 0.67 |
| 16 | [Ne]$3s^23p^4$ [$^3P$] → [Ne]$3s^23p^4$ [$^1D$] | 1.15 | 0.46 | 0.42 | 0.38 |

**Table S1:** Spin-flip excitation energy predictions of LSDA, PBE, and r$^2$SCAN. Column 1 gives atomic number. Column 2 gives electron configuration of ground (left) and excited (right) states electron transitions between with term. Columns 3-6 give excitation energies from experimental values or functional prediction in eV. Experimental energies obtained from NIST.[9]

4. Valence and Rydberg Excitation MAEs in and MAREs as a Subset of Non-Aufbau Excitations

| | LSDA | PBE | r$^2$SCAN |
|---|---|---|---|
| MAE | 0.16 | 0.18 | 0.42 |
| MARE (%) | 3.90 | 4.89 | 9.39 |

**Table S3:** Valence excitation energy prediction MAEs and MAREs of LSDA, PBE, and SCAN. MAEs given in eV.

| | LSDA | PBE | $r^2$SCAN |
|---|---|---|---|
| MAE | 0.21 | 0.12 | 0.16 |
| MARE (%) | 1.90 | 1.17 | 2.10 |

**Table S4:** Rydberg excitation energy prediction MAEs and MAREs of LSDA, PBE, and SCAN. MAEs given in eV.